\documentclass[preprint,showpacs,aps,prc,groupedaddress,floatfix]{revtex4}
\usepackage{epsfig}
\usepackage{amsmath,amssymb}
\usepackage{bm}

\newcommand{\beq}{\begin{equation}}
\newcommand{\eeq}{\end{equation}}

\def\nuc#1#2{\relax\ifmmode{}^{#1}{\protect\text{#2}}\else${}^{#1}$#2\fi}

\begin{document}
\graphicspath{{figures/}}

\title{Diagnosing `Diagnosing the trouble\ldots'}
\author{R. S. Mackintosh}
\email{raymond.mackintosh@open.ac.uk}
\affiliation{School of Physical Sciences, The Open University, Milton Keynes, MK7 6AA, UK}

\date{\today}

\begin{abstract}Stimulated by a recent preprint of Bricmont and Goldstein.

\end{abstract}

\maketitle

Bricmont and Goldstein~\cite{BG}  make the case for the de Broglie Bohm interpretation of quantum mechanics, but I believe that their case has something missing, as I shall try to explain. In his famous 1964 Messenger lectures (see also Ref.~\cite{feynman}), Richard Feynman remarked that the behaviours of  electrons and photons `are both screwy, but in exactly the same way', and that has always struck me as true and important. But photons are studiously avoided in most discussions of  de Broglie Bohm (dBB) QM. Where is the dBB treatment of even the single slit diffraction of light? Where is the dBB treatment of  light emitted in the classic case of atomic hydrogen, the case that set Bohr on the road to QM? Until there is a full dBB treatment of photons, dBB advocates must understand that many will suspect that the dBB formalism is a remarkable mapping, a property of the Schr\"odinger equation, for which there is no equivalent for photons. 

Most dBB discussions apply the  theory to particles interacting with slits, but the world is made of collections of bound particles: atoms, molecules and solids. In particular,  nuclei are studiously avoided. We believe that all \nuc{14}{C} nuclei in the ground state are identical. Being spin zero bosons, two \nuc{14}{C} nuclei, if they could be made to scatter from each other, would show the interference pattern that Mott showed was appropriate for identical bosons. Yet the \nuc{14}{C} nuclei will not decay at identical times. So where is the hidden variable that  will determine when the decay will take place? Presentations of dBB theory always identify the `hidden variable' as the wave function solution of Schr\"odinger's equation, with the position of the particle explicit,  but what nucleons in \nuc{14}{C} have explicit positions?  And how do the hidden variables then convey determined paths to the emitted electron and antineutrino? This seems to be an implication of dBB.  

Of course, weak interaction processes, as they apply to \nuc{14}{C} decay,  may be beyond dBB at the moment, but what about the decay of an excited state of a helium atom? Somehow an excited entangled state of two electrons is going to make a transition to a lower energy entangled state of two electrons, emitting a photon in the process. All members of a collection of helium atoms in some particular excited state are identical, so where is the hidden dBB variable that distinguishes when one particular atom in the collection will decay by emitting a photon? And what hidden variable will determine the direction in which the photon,  emitted as a dipole wave, appears at a detector? The recoiling atom will be entangled with the photon, so it is the centre of mass motion of the photon and atom that presumably is a dBB-type beable. Remember, atoms and photons are screwy in exactly the same way. 

Identical excited helium atoms might be a hard to deal with, so let us consider a collection of identical \nuc{238}{U} nuclei, each in the ground state (as rapidly they all will be). They will decay by emitting an alpha particle in a manner according to the following objective property: they have a propensity for decaying such that a measurement of the half life found for a collection of them would yield a value of $4.468 \times 10^9$ years. That is an objective property, and consideration of the propensity for decay means that realism need not be abandoned, only the nature of the `real' changes. (At least dBB is correct in identifying the problem as ontological.) So where is the hidden variable that determines when a particular \nuc{238}{U}  nucleus decays? Of course, one could apply the dBB formalism in an up-to-date version of the famous Gamow tunnelling calculation. In such a dBB calculation, one would assume that the position of the alpha particle for some particular nucleus would be the local beable. But the case is not that simple since the Gamow picture assumes a preformed alpha particle within the nucleus, and this is not realistic for a complete calculation. An alpha particle needs to emerge from a complex correlated system of identical protons and identical neutrons, so identifying the appropriate beables might not be obvious. If dBB advocates could provide a complete analysis of alpha decay, that  might help persuade people of the dBB picture.

Incidentally, alpha decay provides another window on an all-to-common misinterpretation of QM. Billions of years ago tiny grains of mineral loaded with uranium and thorium got trapped between layers of mica. Over millions or billions of years, alpha particles were emitted as the \nuc{238}{U} or \nuc{232}{Th} nuclei decayed. They discoloured the mica as they came towards the end of their `paths', forming rings in the mica around the grain of mineral, the radii of the rings corresponding to the energy of the emitted alpha particles. The resulting `pleochroic  haloes' were studied by Joly~\cite{Joly} and many others and are illustrated in a book by Rutherford~\cite{Rutherford}.  My point is that those haloes do not come into existence when the mica is placed under a microscope and examined by a human (with a PhD, as Bell might have said). This should be borne in mind by anyone who is tempted to consider that `measurement'  in a QM context has anything to do with `measurement' in the context of human  (or feline) actions. The thing in common is just basically the same set of eleven letters. Will we ever see a full Bohmian description of the production of pleochroic halos?

The motivation for dBB seems to be the restoration of `particleness' to entities we call electrons, protons, silver atoms,  etc. But there is a price to pay:   the un-particle-like trajectories. Maybe the ontology implicit in good old Copenhagen is closer to the truth: a particle is an entity that, when questioned as to location, answers with a single place, or more exactly a single interval $\Delta x$ corresponding to the resolution of the equipment. Of course, that entails non-locality, a matter of deep concern to Einstein in 1927, but proponents of dBB QM cannot criticise that.

\end{document}